\documentclass[12pt]{article}
\usepackage{amsmath,amssymb}
\usepackage{bm}
\usepackage{mathrsfs}
\usepackage{fourier}
\usepackage{multirow}
\allowdisplaybreaks[4]
\newcommand{{\Slashp}}{p\!\!\!\!\!\big/}
\newcommand{{\Slashq}}{q\!\!\!\!\!\big/}
\usepackage[usenames]{color}

\begin{document}

\title{Finiteness, duality, and fermionic symmetry}

\author{
Yoshiharu \textsc{Kawamura}\footnote{E-mail: haru@azusa.shinshu-u.ac.jp}\\
{\it Department of Physics, Shinshu University, }\\
{\it Matsumoto 390-8621, Japan}\\
}

\date{
March 13, 2015}

\maketitle
\begin{abstract}
We propose a framework for a new type of finite field theories 
based on a hidden duality 
between an ultra-violet and an infra-red region.
Physical quantities do not receive radiative corrections
at a fundamental scale or the fixed point of the duality transformation,
and this feature is compatible with models 
possessing a specific fermionic symmetry.
Theories can be tested indirectly by relations among parameters,
reflecting underlying symmetries.
\end{abstract}


\section{Introduction}

Quantum field theory has perplexed physicists 
with the appearance of infinities on radiative corrections
of physical quantities,
but the problem has offered useful hints for a fundamental theory.
In well-behaved theories such as the quantum electrodynamics (QED)
and the standard model (SM),
infinities are removed by the regularization 
and the renormalization procedure.
There, however, exists a non-renormalizable theory 
such as the quantum version of Einstein gravity,
and it suggests the idea
that {\it an underlying theory must own a finiteness}.

Typical examples
are superstring theories (SSTs)~\cite{SST} 
and finite field theories (FFTs)~\cite{FFT}.
The finiteness comes from 
the world-sheet modular invariance in SSTs 
and the vanishing of $\beta$-functions in FFTs.

Both have a feature that a fundamental energy scale $\varLambda$
exists, but its origin is different from each other.
In SSTs, $\varLambda$ is the string scale defined by the string tension.
The world-sheet modular invariance implies 
that radiative corrections from the contributions below $\varLambda$
are equivalent to those above $\varLambda$
and are given by the integration of an independent region
called $\lq\lq$the fundamental region''.
There, $\varLambda$ plays the role of cut-off parameter
or the ultra-violet (UV) divergences are identified with
unphysical infra-red (IR) ones.
In FFTs, the theory becomes scale invariant with the vanishing of
$\beta$-functions at $\varLambda$.
Then, physical parameters do not run beyond the scale, 
and the concept of scale becomes vague. 
Models have a high calculability and predictability,
because the reduction of coupling constants is
realized and the particle contents are tightly restricted,
combined with the grand unification~\cite{FUT}.

Both theories are powerful candidates for the physics at $\varLambda$.
However, any evidences for new physics 
beyond the SM, e.g., supersymmetry (SUSY),
compositeness and extra dimensions, have not been discovered.
Hence, it would be meaningful to pursue other possibilities.

In this paper, we use the secret of finiteness in SSTs
as a guide for constructing a new type of FFTs,
and propose a framework of
theories based on a hidden duality 
between an UV and an IR region.
Physical quantities do not receive radiative corrections
at the fixed point of the duality transformation,
and this feature is compatible with models
possessing a specific fermionic symmetry.
Theories can be tested indirectly by relations among parameters,
reflecting underlying symmetries, 
in case that the system has a large symmetry at a fundamental level.

The contents of this paper are as follows.
We present novel FFTs based on
a duality relating world lines in Sect. II.
We explain that the finiteness can be
assisted by an exotic symmetry concerning abnormal particles
and theories can be tested indirectly in Sect. III.
Section IV is devoted to conclusions and discussions.

\section{Finiteness based on world-line duality}

We take the standpoint that quantum field theory
is an effective description of an unknown underlying theory,
and FFTs can be constructed by bringing in features of
the fundamental theory.

First, we list the assumptions relating features
of the ultimate theory.\footnote{
Our idea is inspired by the world-sheet modular invariance 
in closed string theories,
and the proposal of solving the gauge hierarchy problem
and the cosmological constant problem by Dienes~\cite{Dienes}.
}\\
(a) There is an energy scale $\varLambda$ that associated with
a property of fundamental objects such as the string scale.
The fundamental objects possess various states that
identified with elementary particles.
The ground states are regarded
as massless particles, and some of them acquire small masses
$m_k (\ll \varLambda)$ through a low-energy dynamics.
The excited states are particles with masses of $O(\varLambda)$.\\
(b) There is a duality between the physics
at a higher-energy scale ($\mu \gtrsim \varLambda$) 
and that at a lower-energy scale ($\mu \lesssim \varLambda$).
Physical quantities are invariant under the duality transformation,
and are estimated as finite values using one of the energy regions.\\
(c) A remnant of the duality is hidden in quantities of
the low-energy physics involved with $\varLambda$, 
e.g., radiative corrections on parameters~\cite{duality}.
Finite corrections can be incorporated in
our formulation with a slight modification of 
the duality transformation,
in the presence of low-energy parameters such as $m_k$.

To illustrate our idea, let us consider quantum corrections 
on a quantity $A$ at the one-loop level given by,
\begin{eqnarray}
&~& \delta A(q) = \int_{0}^{1} ds \int_0^{\infty} dt 
\int_{0}^{\infty} \frac{d^4p}{(2\pi)^4}~f(p, q, g_i, m_k, s, t)
\nonumber\\
&~& ~~~~~~~~~~~
= \int_{0}^{1} ds \int_0^{\infty} dt~h(q, g_i, m_k, s, t),
\label{delta-a}
\end{eqnarray}
where $s$ is an integration variable, 
$t$ is the integration variable called a $\lq\lq$proper time'',
$p$ is an Euclidean momentum of a particle running in a loop, 
$q$ is an Euclidean momentum of a particle for an external line,
$g_i$ are coupling constants,
and $h(q, g_i, m_k, s, t)$ is a function of $q$, $g_i$, $m_k$, $s$ and $t$.

In case that $\delta A(q)$ diverges, the infinities come from 
$t = 0$ (corresponding the UV divergences)
and/or $t = \infty$ (corresponding the IR ones),
it can be regularized as
\begin{eqnarray}
\delta A(q)_{\rm reg} = \int_{0}^{1} ds 
\int_{1/\tilde{\varLambda}^2}^{1/\tilde{\mu}^2} dt~h(q, g_i, m_k, s, t),
\label{delta-a-reg}
\end{eqnarray}
where $\tilde{\varLambda} = \tilde{\varLambda}(\varLambda, s)$.
In most cases, $\tilde{\varLambda} = x(s) \varLambda$
and $\tilde{\mu} = y(s) q^2 + y_k(s) m_k^2$
where $x(s)$, $y(s)$ and $y_k(s)$ are functions of $s$.
Although these functions can vanish at some values of $s$,
we often substitute $\tilde{\varLambda}$ and $\tilde{\mu}$
for $\varLambda$ and $q$, respectively.
For instance, $\tilde{\varLambda}$ and $\tilde{\mu}$
are given by $\tilde{\varLambda}^2 = s \varLambda^2$
and $\tilde{\mu}^2 = s(1-s) q^2 + s \mu_{\gamma}^2 + (1-s) m_{\rm e}^2$
for the self-energy of electron in QED~\cite{duality}.
Here, $\mu_{\gamma}$ is a fictitious photon mass
for a regularization of IR divergences
and $m_{\rm e}$ is the electron mass.

We require that $\delta A(q)_{\rm reg}$ should be invariant 
under the following duality transformation on the internal world-line,
\begin{eqnarray}
t \to \frac{1}{\tilde{\varLambda}^4 t}.
\label{t-transf}
\end{eqnarray}
Under the transformation (\ref{t-transf}), 
$\delta A(q)_{\rm reg}$ transforms as
\begin{eqnarray}
\delta A(q)_{\rm reg} \to
\int_{0}^{1} ds 
\int_{\tilde{\mu}^2/\tilde{\varLambda}^4}^{1/\tilde{\varLambda}^2} 
dt~\frac{h(q, g_i, m_k, s, 1/(\tilde{\varLambda}^4 t))}{\tilde{\varLambda}^4 t^2}.
\label{delta-a-reg-t-transf}
\end{eqnarray}
From the equality of the right-hand side of
(\ref{delta-a-reg}) and that of (\ref{delta-a-reg-t-transf}),
the form of $h(q, g_i, m_k, s, t)$ is fixed as 
$h(q, g_i, m_k, s, t)=c_{-1}(q, g_i, m_k, s)/t$ 
and then $\delta A(q)_{\rm reg}$ is determined as
\begin{eqnarray}
\delta A(q)_{\rm reg} = \int_0^1 ds~
c_{-1}(q, g_i, m_k, s) \ln \frac{\tilde{\varLambda}^2}{\tilde{\mu}^2}.
\label{delta-a-reg-det}
\end{eqnarray}
The $\delta A(q)_{\rm reg}$ contains
a $\varLambda$-dependent logarithmic part and finite corrections.\footnote{
It is not clear whether all corrections including higher loops 
are obtained in the duality invariant form.
If not, such corrections can be regarded as a tiny violation of
the duality.
}
We find that quantum corrections 
corresponding quadratic divergences are removed, 
and hence the naturalness problem in the SM
(the quadratic divergence problem relating the Higgs boson mass)
can be solved~\cite{duality}.

So far, we study corrections at the lower energy region 
($q^2 < \varLambda^2$) by considering
only contributions from the ground states 
(corresponding massless particles at $\varLambda$).
Beyond $\varLambda$, threshold corrections 
due to excited states (corresponding massive particles of $O(\varLambda)$)
appear, and following two problems occur.\\
(i) To examine whether physical quantities 
are formulated in a manner consistent with the duality,
{\it for all energy regions}.\\
(ii) To examine whether an effective field theory 
have a high calculability and predictability,
{\it without knowing full spectrum of excited states}.

The threshold corrections might be introduced,
in the form that $h$ contains $\varLambda$ such that
\begin{eqnarray}
h(q, g_i, m_k, s, t)=
\frac{\tilde{c}_{-1}(q, g_i, m_k, s, t+1/(\tilde{\varLambda}^4 t))}{t}.
\label{h-Lambda}
\end{eqnarray}
However, we have no method to determine 
the form of $\tilde{c}_{-1}$ as a function of $t$, 
in the framework of effective field theory, and hence
we pursue an alternative.

Here, we give a bold conjecture that 
{\it there are no threshold corrections due to excited states}.
If it holds, the second problem can be solved,
as will be explained in the next section.

For the first problem,
let us make an adventurous attempt
that the duality can be also applied to the external lines.
We require that $\delta A(q)_{\rm reg}$ should be invariant 
under the duality transformation
$q \to q'$ that corresponds to
\begin{eqnarray}
\tilde{\mu}^2 \to 
\tilde{\mu}'^2 = \frac{\tilde{\varLambda}^4}{\tilde{\mu}^2}.
\label{mu-transf}
\end{eqnarray}
The invariance is written as
\begin{eqnarray}
\delta A(q)_{\rm reg} = \delta A(q')_{\rm reg}.
\label{delta-a-reg-duality}
\end{eqnarray}

From (\ref{delta-a-reg-det}), (\ref{mu-transf}) and (\ref{delta-a-reg-duality}),
we obtain the expressions,
\begin{eqnarray}
\delta A(q)_{\rm reg} 
= \int_{0}^{1} ds \int_{1/\tilde{\varLambda}^2}^{1/\tilde{\mu}^2} dt~
h(q, g_i, m_k, s, t)~~~~
({\rm for}~~\tilde{\mu} \le \tilde{\varLambda})
\label{delta-a-reg-mu}
\end{eqnarray}
and
\begin{eqnarray}
\delta A(q')_{\rm reg} 
= \int_{0}^{1} ds \int_{1/\tilde{\mu}'^2}^{1/\tilde{\varLambda}^2} dt~ 
h(q', g_i, m_k, s, t)~~~~
({\rm for}~~\tilde{\mu}' \ge \tilde{\varLambda}).
\label{delta-a-reg-mu'}
\end{eqnarray}
Under the transformation (\ref{t-transf}),
$\delta A(q)_{\rm reg}$ transforms into 
$\delta A(q')_{\rm reg}$ and vice versa,
and they take a same value with $h(q, g_i, m_k, s, t)=c_{-1}(q, g_i, m_k, s)/t$.
The equality (\ref{delta-a-reg-duality}) comes from
the fact that the distance between 
$1/\tilde{\varLambda}^2$ and $1/\tilde{\mu}^2$ equals to 
that between $1/\tilde{\mu}'^2 (=\tilde{\mu}^2/\tilde{\varLambda}^4)$ 
and $1/\tilde{\varLambda}^2$
at the logarithmic scale.

We find that $\tilde{\varLambda}$ is the fixed point 
under the transformation (\ref{mu-transf}).
The relation (\ref{delta-a-reg-mu'}) suggests a strange feature
that the value obtained by integrating out the degrees
from $\tilde{\varLambda}$ to $\tilde{\mu}' (> \tilde{\varLambda})$
is not the value at $\tilde{\varLambda}$
but that at $\tilde{\mu}'$ (or $q'$),
and the quantity does not receive radiative corrections
at $\tilde{\varLambda}$, i.e., 
$\delta A(q)_{\rm reg}|_{\tilde{\mu} =\tilde{\varLambda}} = 0$.
We give a speculation of such an opposite running of physical quantities
beyond $\tilde{\varLambda}$.
If the role of energy and momentum is exchanged
in the region beyond $\tilde{\varLambda}$,
as is the case that the role of time and space is exchanged inside a black hole,
$p^2$ and $m_k^2$ in $f(p, q, g_i, m_k, s, t)$ can change its sign.
Eventually, it can induce the exchange of integration region.
Then, $\varLambda$ might be the Planck scale $M_{\rm Pl}$.\footnote{
As another work to show the importance of the trans-Planckian physics,
Volovik gave the observation that the sub-Planckian and
trans-Planckian contributions to the vacuum energy
are canceled by the thermodynamical argument~\cite{Volovik}.
}

Our procedure is regarded as not a mere regularization 
but a recipe to obtain finite physical values, 
because $\tilde{\varLambda}$ is (big but) finite and
infinities are taken away by the symmetry relating integration variables, 
like closed string theories.
It is also regarded as the operation to pick out duality invariant parts.
In case that $h(q, g_i, m_k, s, t)$ does not contain $\varLambda$,
it is simply denoted by
\begin{eqnarray}
&~& \delta A(q)_{\rm reg}  
= {\rm Du}\left[\int_{0}^{1} ds \int_0^{\infty} dt~h(q, g_i, m_k, s, t)\right] 
\nonumber \\
&~& ~~~~~~~~~~~~~~~~~ 
= {\rm Du}\left[\int_{0}^{1} ds \int_0^{\infty} dt~
\sum_n c_n(q, g_i, m_k, s) t^n\right]
\nonumber \\
&~& ~~~~~~~~~~~~~~~~~ 
= c_{-1}(q, g_i, m_k, s) \ln \frac{\tilde{\varLambda}^2}{\tilde{\mu}^2},
\label{delta-a-Du}
\end{eqnarray}
where ${\rm Du}[*]$ represents the operation, 
and $h(q, g_i, m_k, s, t)$ is expanded in a series of $t$.

\section{Fermionic symmetry}

\subsection{Calculability}

Let us construct a theory with a high calculability
\footnote{
In this paper, a $\lq\lq$calculability'' means that
physical quantities can be calculated 
in terms of free parameters by a theory,
and a $\lq\lq$predictability'' means that
some features beyond the theory,
such as relations among parameters, can be predicted.
},
based on the feature 
that any threshold corrections do not appear around $\varLambda$.

We assume that fundamental objects have a specific fermionic symmetry
and most states become unphysical by a counterpart of
the quartet mechanism~\cite{K&O1,K&O2}
in a system with the BRST symmetry.

Concretely, all excited states form the quartets such as 
($\varphi_a, c_a, \overline{c}_a, \overline{\varphi}_a$),
and they transform as
\begin{eqnarray}
\bm{\delta}_{\rm f} \varphi_a = (\pm) c_a,~~
\bm{\delta}_{\rm f} c_a = 0,~~
\bm{\delta}_{\rm f} \overline{c}_a = (\pm) \overline{\varphi}_a,~~
\bm{\delta}_{\rm f} \overline{\varphi}_a = 0, 
\label{delta-f}
\end{eqnarray} 
where $(\pm)$ represents $+1$ or $-1$.
The transformation is generated 
by a fermionic conserved charge $Q_{\rm f}$
with the nilpotency, i.e., ${Q_{\rm f}}^2 = 0$.
If we impose suitable subsidiary conditions 
containing the following one on states
in order to select physical states,
\begin{eqnarray}
Q_{\rm f} |{\rm phys}\rangle =0,
\label{phys}
\end{eqnarray} 
all $Q_{\rm f}$-quartets (or two sets of $Q_{\rm f}$-doublets) 
become unphysical.
Hereafter, we denote the set of $Q_{\rm f}$-quartets
as $(\{\varphi_{\rm q}\}, \{c_{\rm q}\})$
and refer to particles belonging in $\{c_{\rm q}\}$ as ghosts.

As a possible candidate of fermionic symmetry,
symmetries between ordinary particles 
and their ghost counterparts
obeying opposite statistics have been proposed~\cite{YK1,YK2,YK3,YK4}.
In this case, $\overline{c}_a$ and $\overline{\varphi}_a$
are the hermitian conjugates of $c_a$ and $\varphi_a$, respectively.
The algebraic relations among relevant conserved charges are given by
\begin{eqnarray}
{Q_{\rm F}}^2 = 0, ~~{Q_{\rm F}^{\dagger}}^2 = 0,~~
\{Q_{\rm F}, Q_{\rm F}^{\dagger}\} = N_{\rm D},
\label{QF-rels}
\end{eqnarray} 
where $Q_{\rm F}$ and $Q_{\rm F}^{\dagger}$ are
a fermionic charge and its hermitian conjugate,
respectively,
and $N_{\rm D}$ is an $U(1)$ charge.
The physical state conditions are imposed as
\begin{eqnarray}
Q_{\rm F} |{\rm phys}\rangle =0,~~
Q_{\rm F}^{\dagger} |{\rm phys}\rangle =0,~~
N_{\rm D} |{\rm phys}\rangle =0.
\label{phys-QF}
\end{eqnarray} 

In case that parts of massless states are $Q_{\rm f}$-singlets
(whose set is denoted by $\{\varphi_{\rm s}\}$)
and are physical,
the system can be, in general, described by 
\begin{eqnarray}
\mathcal{L}_{\rm total} = \mathcal{L}_{\rm s}(\{\varphi_{\rm s}\})
+ \mathcal{L}_{\rm q}(\{\varphi_{\rm q}\}, \{c_{\rm q}\})
+ \mathcal{L}_{\rm mix}(\{\varphi_{\rm s}\}, 
\{\varphi_{\rm q}\}, \{c_{\rm q}\})
= \mathcal{L}_{\rm s}(\{\varphi_{\rm s}\}) + \bm{\delta}_{\rm f} \mathcal{R},
\label{L}
\end{eqnarray} 
where $\mathcal{L}_{\rm s}$, $\mathcal{L}_{\rm q}$ 
and $\mathcal{L}_{\rm mix}$
are the Lagrangian density for $Q_{\rm f}$-singlets, $Q_{\rm f}$-quartets
and interactions between $Q_{\rm f}$-singlets and $Q_{\rm f}$-quartets.
Under suitable subsidiary conditions 
including $Q_{\rm f} |{\rm phys}\rangle =0$ on states,
all $Q_{\rm f}$-quartets become unphysical
and would not give any physical effects on $Q_{\rm f}$-singlets,
that is, $Q_{\rm f}$-singlets do not receive any radiative corrections 
from $Q_{\rm f}$-quartets.
Hence, the theory is free from 
the gauge hierarchy problem~\cite{YK1}.\footnote{
Other type of fermionic symmetry called $\lq\lq$misaligned 
supersymmetry'' has been proposed to solve the gauge hierarchy
problem and to realize the finiteness,
in the absence of space-time SUSY~\cite{Dienes,Dienes2,DM&M}.
}

Let us discuss the calculability in our formulation.
If all massive modes with masses of $O(\varLambda)$
are $Q_{\rm f}$-quartets and unphysical,
there are no threshold corrections at $\varLambda$
and all physical quantities can be calculable
using $\mathcal{L}_{\rm s}(\{\varphi_{\rm s}\})$ alone.
That is, if values of coupling constants $g_i$ and masses $m_k$
are determined by precision measurements,
we can obtain values of physical quantities accurately.
Note that $g_i$ and $m_k$ are free parameters 
and their values are not determined theoretically, 
in the framework of a low-energy effective theory.

Ordinarily, non-renormalizable interactions are
generated as a result that heavy particles are integrated out.
In our system, heavy particles with masses of $O(\varLambda)$
appear by pairs in the interaction terms
because of the $Q_{\rm f}$ invariance.
In the process with ordinary physical particles alone
in the external lines,
heavy particles appear in loops
and the sum of contributions can be canceled out
by the fermionic symmetry.
That is, non-renormalizable interactions
are not induced due to the excited states at $O(\varLambda)$.
Then, there is a possibility that
$\mathcal{L}_{\rm s}(\{\varphi_{\rm s}\})$
is renormalizable at $\varLambda$,
neglecting the effect of gravity.
In this case,
we have an interesting expectation
that {\it the system of visible fields 
is described by the Lagrangian density
containing renormalizable terms alone}.
Because visible fields come from the massless states at $\varLambda$,
the effective Lagrangian density must have symmetries
such as gauge symmetry, chiral symmetry
and/or conformal symmetry
to guarantee the masslessness.

After integrating out extra particles in $\{\varphi_{\rm s}\}$
other than the SM ones,
we arrive at the system described by
\begin{eqnarray}
\mathcal{L}_{\rm SM} + \varDelta\mathcal{L}_{\rm SM},
\label{L-SM}
\end{eqnarray} 
where $\mathcal{L}_{\rm SM}$ 
stands for the renormalizable Lagrangian density for the SM particles
and and $\varDelta\mathcal{L}_{\rm SM}$
is the non-renormalizable one generated by
contributions from heavy particles in $\{\varphi_{\rm s}\}$
beyond the weak scale.
A dark matter and massive right-handed neutrinos
are included as candidates of extra ones.
The exploration of $\varDelta\mathcal{L}_{\rm SM}$
is important to probe the physics (at the terascale) beyond the SM
and to determine $\mathcal{L}_{\rm s}(\{\varphi_{\rm s}\})$
indirectly.

The system above $\varLambda$ can be also described 
by $\mathcal{L}_{\rm total}$
(essentially $\mathcal{L}_{\rm s}(\{\varphi_{\rm s}\})$),
and has a grounding that the duality introduced in the previous section
holds for physical quantities.

\subsection{Predictability}

Next, we discuss the predictability in our formulation
based on $\mathcal{L}_{\rm total}$.
It is needed to specify a structure of system at $\varLambda$.

Let us take a reasonable conjecture that
{\it fundamental objects have a large symmetry intrinsically
and the symmetry is realized in an unbroken phase at $\varLambda$}.
Concretely, the system has the symmetry 
whose transformation group is $G_{\rm U}$ in the unbroken phase.
By some mechanism, $G_{\rm U}$ is broken down to the subgroup $G$
and the system is described by $\mathcal{L}_{\rm total}$,
where $\{\varphi_{\rm s}\}$ are multiplets of $G$.

Here, we consider two scenarios
that $\mathcal{L}_{\rm total}$ is derived after the reduction of symmetry,
and find a prediction based on them.\\
($\alpha$) First one is that all particles belong to 
the members of $Q_{\rm f}$-quartets 
and are multiplets of $G_{\rm U}$, in the unbroken phase.
They are denoted by $(\{\varphi_{\rm q}^{\rm U}\}, \{c_{\rm q}^{\rm U}\})$.
The system is described by
\begin{eqnarray}
\mathcal{L}^{(\alpha)} 
= \mathcal{L}_{\rm U}^{(\alpha)}(\{\varphi_{\rm q}^{\rm U}\})
+ \mathcal{L}_{\rm gh}^{(\alpha)}(\{c_{\rm q}^{\rm U}\}) 
+ \mathcal{L}_{\rm int}^{(\alpha)}(\{\varphi_{\rm q}^{\rm U}\}, \{c_{\rm q}^{\rm U}\})
= \bm{\delta}_{\rm f} \mathcal{R}_{\rm U}^{(\alpha)},
\label{L-alpha}
\end{eqnarray} 
where $\mathcal{L}_{\rm U}^{(\alpha)}$, $\mathcal{L}_{\rm gh}^{(\alpha)}$ 
and $\mathcal{L}_{\rm int}^{(\alpha)}$
are the Lagrangian densities for ordinary particles 
$\{\varphi_{\rm q}^{\rm U}\}$, 
the ghost counterparts $\{c_{\rm q}^{\rm U}\}$,
and interactions between ordinary particles and ghosts.

The multiplets of $G_{\rm U}$ are decomposed into those of $G$ such that
\begin{eqnarray}
\{\varphi_{\rm q}^{\rm U}\} 
\Rightarrow \{\varphi_{\rm q}\}_0 + \{\varphi_{\rm q}\}_1,~~
\{c_{\rm q}^{\rm U}\} 
\Rightarrow \{c_{\rm q}\}_0 + \{c_{\rm q}\}_1.
\label{G-decomp}
\end{eqnarray} 
If some ghosts $\{c_{\rm q}\}_0$ disappear,
ordinary particles $\{\varphi_{\rm q}\}_0$ 
turn out to be $Q_{\rm f}$-singlets
and the reduction of symmetry occurs.\footnote{
It has been reported that,
in a system with complex scalar fields on a higher-dimensional space-time,
some physical modes are released from 
unphysical $Q_{\rm F}$-doublets
and the reduction of a large symmetry occurs
through the orbifold breaking mechanism~\cite{YK3}.
}
Then, the system is described by
\begin{eqnarray}
{\mathcal{L}'}^{(\alpha)} 
= \mathcal{L}_{\rm U}^{(\alpha)}(\{\varphi_{\rm s}\}, \{\varphi_{\rm q}\})
+ {\mathcal{L}'}_{\rm gh}^{(\alpha)}(\{c_{\rm q}\}) 
+ {\mathcal{L}'}_{\rm int}^{(\alpha)}(\{\varphi_{\rm s}\}, \{\varphi_{\rm q}\}, \{c_{\rm q}\}),
\label{L'-alpha}
\end{eqnarray}
where $\{\varphi_{\rm q}\}_0$, $\{\varphi_{\rm q}\}_1$ 
and $\{c_{\rm q}\}_1$
are denoted by $\{\varphi_{\rm s}\}$, $\{\varphi_{\rm q}\}$ 
and $\{c_{\rm q}\}$.
In this scenario, ${\mathcal{L}'}^{(\alpha)}$ corresponds to 
$\mathcal{L}_{\rm total}$ at $\varLambda$.\\
($\beta$) Second one is that some massless ordinary particles 
belong to members of $Q_{\rm f}$-singlets $\{\varphi_{\rm s}^{\rm U}\}$ 
and others are $Q_{\rm f}$-quartets 
$(\{{\varphi'}_{\rm q}^{\rm U}\}, \{{c'}_{\rm q}^{\rm U}\})$.
All particles form multiplets of $G_{\rm U}$,
in the unbroken phase.
The system is described by
\begin{eqnarray}
&~& \mathcal{L}^{(\beta)} 
= \mathcal{L}_{\rm U}^{(\beta)}(\{\varphi_{\rm s}^{\rm U}\})
+ \mathcal{L}_{\rm q}^{(\beta)}(\{{\varphi'}_{\rm q}^{\rm U}\}, \{{c'}_{\rm q}^{\rm U}\}) 
+ \mathcal{L}_{\rm int}^{(\beta)}(\{\varphi_{\rm s}^{\rm U}\}, 
\{{\varphi'}_{\rm q}^{\rm U}\}, \{{c'}_{\rm q}^{\rm U}\})
\nonumber \\
&~& ~~~~~~~~~
= \mathcal{L}_{\rm U}^{(\beta)}(\{\varphi_{\rm s}^{\rm U}\}) 
+ \bm{\delta}_{\rm f} \mathcal{R}_{\rm U}^{(\beta)},
\label{L-beta}
\end{eqnarray} 
where $\mathcal{L}_{\rm U}^{(\beta)}$, $\mathcal{L}_{\rm q}^{(\beta)}$ 
and $\mathcal{L}_{\rm int}^{(\beta)}$
are the Lagrangian densities for $\{\varphi_{\rm s}^{\rm U}\}$, 
$Q_{\rm f}$-quartets, 
and interactions among them.
The $Q_{\rm f}$-singlets are decomposed into those of $G$ such that
\begin{eqnarray}
\{\varphi_{\rm s}^{\rm U}\} 
\Rightarrow \{\varphi_{\rm s}\}_0 + \{\varphi_{\rm s}\}_1.
\label{G-decomp-beta}
\end{eqnarray} 
If some ghosts $\{c_{\rm s}\}_1$ appear and
they form $Q_{\rm f}$-quartets in company with $\{\varphi_{\rm s}\}_1$,
the reduction of symmetry occurs.
Then, the system is described by
\begin{eqnarray}
&~& {\mathcal{L}'}^{(\beta)} 
= \mathcal{L}_{\rm U}^{(\beta)}(\{\varphi_{\rm s}\}, \{\varphi_{\rm q}\}_1)
+ {\mathcal{L}'}_{\rm gh}^{(\beta)}(\{c_{\rm q}\}_1) 
+ \mathcal{L}_{\rm q}^{(\beta)}(\{{\varphi'}_{\rm q}^{\rm U}\}, \{{c'}_{\rm q}^{\rm U}\}) 
\nonumber \\
&~& ~~~~~~~~~~~~~
+ {\mathcal{L}'}_{\rm int}^{(\beta)}(\{\varphi_{\rm s}\}, \{\varphi_{\rm q}\}_1,
\{c_{\rm q}\}_1, \{{\varphi'}_{\rm q}^{\rm U}\}, \{{c'}_{\rm q}^{\rm U}\}),
\label{L'-beta}
\end{eqnarray}
where $\{\varphi_{\rm s}\}_0$, $\{\varphi_{\rm s}\}_1$ 
and $\{c_{\rm s}\}_1$
are denoted by $\{\varphi_{\rm s}\}$,
$\{\varphi_{\rm q}\}_1$ and $\{c_{\rm q}\}_1$,
and $\{{\varphi'}_{\rm q}^{\rm U}\}$ and $\{{c'}_{\rm q}^{\rm U}\}$ 
are also used
in place of their decompositions under $G$, for simplicity. 
In this scenario, ${\mathcal{L}'}^{(\beta)}$ corresponds to 
$\mathcal{L}_{\rm total}$ at $\varLambda$.\footnote{
There is a possibility that $\mathcal{L}_{\rm total}$ is also derived
after eliminating some parts ($\{\varphi_{\rm s}\}_1$)
of $Q_{\rm f}$-singlets,
starting from $\mathcal{L}^{(\beta)}$.
}

Unless extra contributions appear on the symmetry reduction,
${\mathcal{L}'}^{(\alpha)}$ and ${\mathcal{L}'}^{(\beta)}$ 
must fit closely with 
$\mathcal{L}_{\rm total}$ at $\varLambda$ from the matching condition.
Because both 
$\mathcal{L}_{\rm U}^{(\alpha)}(\{\varphi_{\rm s}\}, \{\varphi_{\rm q}\})$
and $\mathcal{L}_{\rm U}^{(\beta)}(\{\varphi_{\rm s}\}, 
\{\varphi_{\rm q}\}_1)$
respect the large symmetry $G_{\rm U}$,
we expect that specific relations among some parameters in
$\mathcal{L}_{\rm s}(\{\varphi_{\rm s}\})$
hold at the tree level such that
\begin{eqnarray}
\left. g_1 = g_2 = \cdots = g_l \right|_{\varLambda},
\label{g-rel}
\end{eqnarray} 
reflecting the underlying symmetry.
Note that the symmetry of the system enhances
from $G$ to $G_{\rm U}$
in the limit that all ghost fields go to zero,
because other terms than 
${\mathcal{L}}_{\rm U}^{(\alpha)}$ 
and ${\mathcal{L}}_{\rm U}^{(\beta)}$ contain ghosts.

The relations (\ref{g-rel}) hold at the quantum level
(up to some gravitational effects),
because parameters do not receive threshold corrections 
at $\varLambda$ in our theory.
Then, (\ref{g-rel}) can become fingerprints or indirect proof
of the underlying symmetry and the existence of unphysical fields,
if gravitational effects are small enough or predictable.
We can test them, using the measured values for $g_k$ 
and the renormalization group equations, 
based on $\mathcal{L}_{\rm s}(\{\varphi_{\rm s}\})$.
Note that (\ref{g-rel}) are not understood from $\mathcal{L}_{\rm s}$ alone,
because the symmetry of $\mathcal{L}_{\rm s}$ 
is not $G_{\rm U}$ but $G$,
as seen from the fact that
$\{\varphi_{\rm s}\}$ form not multiplets of $G_{\rm U}$
but those of $G$.

If fundamental objects have a large gauge symmetry to unify
the SM gauge group $SU(3)_{\rm C} \times SU(2)_{\rm L} \times U(1)_{\rm Y}$,
we predict the unification of gauge coupling constants~\cite{YK1}.
This prediction suggests the existence of extra particles 
other than the SM ones beyond the weak scale.\footnote{
For instance, the unification of gauge coupling constants at $M_{\rm Pl}$
has been studied under the assumption
that extra particles appear at the terascale
as remnants of hypermultiplets~\cite{YK5}.
}

\section{Conclusions and discussions}

We have used the secret of finiteness in SSTs
as a guide for constructing a new type of FFTs,
and proposed a framework of
theories based on a hidden duality 
between an UV and an IR region.
Concretely, quantum corrections are
invariant under the duality transformation 
$t \to 1/(\tilde{\varLambda}^4 t)$,
where $t$ is a proper time for a particle running in the loop
and $\tilde{\varLambda}$ is a fundamental scale $\varLambda$
up to some factor.
From the requirement that radiative corrections should be also
invariant under the duality transformation
for external world lines,
we have arrived at the feature that
physical quantities cannot receive any radiative corrections
at $\tilde{\varLambda}$, neglecting the gravitational effects.
It is compatible with models possessing a specific fermionic symmetry.
The low-energy physics $(q^2 \ll \tilde{\varLambda}^2)$ can be described by
a renormalizable effective field theory 
whose constituents are $Q_{\rm f}$-singlets,
and physical quantities can be calculated precisely
using the low-energy theory alone.
Theories can be tested indirectly by relations among parameters,
reflecting underlying symmetries, in case that the system
has a large symmetry at the ultimate level.

Here, let us summarize features of fundamental theory, in the broken phase.
The theory is defined just at the fundamental scale.
Values of fundamental parameters are given there
as initial conditions.
The sector with ordinary particles alone has the large symmetry
$G_{\rm U}$,
that can originate from a characteristics of fundamental objects.
Ordinary particles are decomposed into multiplets of the subgroup $G$.
Some submultiplets are observed, 
and others form $Q_{\rm f}$-quartets
and become unphysical in the presence of ghosts.
The reduction of symmetry occurs from $G_{\rm U}$ to $G$,
as a result that ghosts are not multiplets of $G_{\rm U}$ 
but those of $G$.
The symmetry of the system enhances
in the limit that all ghost fields go to zero.

There are serious problems related to ghost fields.
First, in most cases, ghosts require the non-local interactions~\cite{YK1}, 
and this fact might suggest that
fundamental objects are not point particles but
extended objects, and
a formulation using extended objects should be required
to describe interactions containing ghosts consistently.
Second, the origin of incomplete multiplets for ghosts is not clear.
We have introduced two scenarios.
One is that some ghosts are reduced
for the system that
all members are $Q_{\rm f}$-quartets as a beginning.
The other is that ghosts emerge for the system that
some ordinary particles exist as $Q_{\rm f}$-singlets.
In this case, ghosts might appear as solitonic states.
In any case,
both scenarios seem to be unrealistic or surreal, 
because a drastic change for the degrees of freedom
is needed to realize them.
Though we recognize that quantum field theory is an extremely useful tool
to describe the lower-energy dynamics than $\varLambda$,
there is a possibility that it has limitations, i.e., 
every physical phenomenon at $\varLambda$
cannot be described completely in the framework of field theory
with finite numbers of fields.
It is less unnatural that the number of particle species changes
for the system with infinite kinds of fields 
than for that with finite kinds of ones.
Hence, it would be interesting to study the reduction
or emergence of ghosts,
using a theory of extended objects.

\section*{Acknowledgments}
The author thanks Prof. G. Zoupanos 
for valuable discussions and useful information
of finite field theories.
This work was supported in part by scientific grants 
from the Ministry of Education, Culture,
Sports, Science and Technology under Grant No.~22540272.

\end{document}